\documentclass[twocolumn,superscriptaddress,amsfont,amssymb,amsmath, showpacs,balancelastpage]{revtex4-1}
\usepackage{CJKutf8}

\usepackage{amsthm}
\newtheorem{theorem}{Theorem}

\newtheorem{proposition}[theorem]{Proposition}

\usepackage[utf8]{inputenc}
\usepackage{amsmath,amssymb,amsfonts,stmaryrd}
\usepackage{physics}
\usepackage{dsfont}
\usepackage{graphicx}
\usepackage{xcolor}
\usepackage{graphicx}
\usepackage{cancel}
\usepackage{natbib}
\usepackage{color}
\usepackage{tikz}
\usepackage{tikz-cd}
\usepackage{mathrsfs}
\usepackage{relsize}
\usepackage{multirow}
\usepackage[linktocpage]{hyperref}
\usepackage[all]{xy}
\hypersetup{colorlinks=true,citecolor=blue,linkcolor=blue, urlcolor=blue, breaklinks=true}

\usepackage{float}

\usepackage{bm} 
\usepackage{subfigure} 
\usepackage{environ}
\usepackage{url}
\usepackage[margin=0.75in]{geometry}
\usepackage{ulem}

\usepackage{mathtools}

\newcommand{\cA}{{\mathcal A}}

\newcommand{\cT}{{\mathcal T}}
\newcommand{\Mod}{\mathbf{Mod}}
\newcommand{\fd}{\mathrm{fd}}
\newcommand{\bTheta}{\mathbf{\Theta}}
\newcommand{\cF}{\mathcal{F}}
\newcommand{\cB}{\mathcal{B}}

\newcommand{\cE}{\mathcal{E}}
\newcommand{\cM}{\mathcal{M}}
\newcommand{\Fun}{\mathrm{Fun}}

\newcommand{\Z}{{\mathbb Z}}

\NewEnviron{eqs}{%
\begin{equation}\begin{split}
    \BODY
\end{split}\end{equation}}

\definecolor{dkgreen}{rgb}{0,0.5,0}

\newcommand{\change}[1]{\textcolor{black}{#1}}

\theoremstyle{definition}
\newtheorem{definition}{Definition}

\theoremstyle{remark}

\newtheorem*{example}{Example}

\begin{document}

\begin{CJK*}{UTF8}{bsmi}

\title{Sixteen-Fold Way for Fermionic Topological Orders}

\author{Ryohei Kobayashi}
\email[E-mail: ]{ryohei.k@ap.t.u-tokyo.ac.jp}
\affiliation{Department of Physics, The University of Tokyo, 7-3-1 Hongo, Bunkyo-ku, Tokyo 113-0033, Japan}

\author{Abhinav Prem}
\email[E-mail: ]{aprem@bard.edu}
\affiliation{Physics Program, Bard College, 30 Campus Road, Annandale-on-Hudson, New York 12504, USA}

\author{Matthew Yu}
\email[E-mail: ]{yumatthew70@gmail.com}
\affiliation{Mathematical Institute, University of Oxford, OX2 6GG, Oxford, UK}

\date{\today}
\begin{abstract}

Fermionic topological orders can host ’t Hooft anomalies with no bosonic counterpart. We identify a new sixteen-fold family of (2+1)D fermionic topological orders, forming a fermionic analogue of Kitaev’s sixteen-fold way. This family is distinguished by the mod 16 ’t Hooft anomaly of a $\mathbb{Z}_2$ one-form symmetry, generated in each theory by a single nontrivial $\mathbb{Z}_2$ anyon. This intrinsically fermionic anomaly permits anyon spins that are forbidden in bosonic phases; the simplest new example is an Abelian fermionic topological order containing a single $\mathbb{Z}_2$ Abelian anyon of spin 1/8. Each theory can be realized as the gapped boundary of a (3+1)D fermionic symmetry-protected topological (SPT) phase protected by the $\mathbb{Z}_2$ one-form symmetry, which acquires a $\mathbb{Z}_{16}$ classification once the spacetime spin structure is twisted by the one-form symmetry. We realize these phases microscopically via lattice models built from Walker--Wang models coupled to local fermions.

 \end{abstract}

\maketitle

\end{CJK*}




\paragraph*{Introduction.}

Symmetry is among the most powerful organizing principles for quantum many-body systems. For ordinary global symmetries, the classification of symmetry-protected topological (SPT) phases and their anomalous boundaries has generated a comprehensive catalogue of phases of matter across condensed matter and quantum field theory~\cite{Chen:2011pg}. A natural extension is provided by higher-form symmetries, which act on extended rather than pointlike objects and are ubiquitous in topological phases; for instance, in $(2+1)$ dimensions, Abelian anyons of a topological order generate 1-form symmetries, and the braiding and spins of those anyons encode the symmetries' 't~Hooft anomalies~\cite{barkeshli2019,Gaiotto:2014kfa}.

In bosonic topological orders (bTOs), the dictionary between anomalies of higher-form symmetries and anyon spins is well-established: for example, a $\mathbb{Z}_2$ 1-form symmetry in (2+1)D is generated by an Abelian anyon whose spin is quantized to $0,\tfrac14,\tfrac12,\tfrac34$, realizing a mod-$4$ anomaly. Fermionic systems allow a much richer possibility of 't Hooft anomalies due to an interplay with fermion parity; such intrinsically fermionic anomalies are now familiar for ordinary ($0$-form) symmetries~\cite{FidkowskiKitaev1d, gu2014, WangGu2020, Kapustin2015spinbordism,barkeshli2021invertible,bhardwajGaiottoKapustin2017, Chen2014TI, Witten:2016cio,Kobayashi2019gapped, Tata2022anomalies, Tong_2020, Hason_2020}. It is hence natural to ask whether higher-form symmetries of fermionic systems can likewise carry anomalies with no bosonic counterpart, and whether these permit anyons whose spin and statistics are forbidden in any bosonic phase.

In this Letter, we answer this affirmatively through a particularly simple construction. We identify a \textit{new sixteen-fold family} of (2+1)D fermionic topological orders (fTOs) -- a fermionic analogue of Kitaev's sixteen-fold way of bTOs~\cite{Kitaev_2006} -- distinguished by the mod-$16$ 't~Hooft anomaly of a $\mathbb{Z}_2$ 1-form symmetry generated, in each theory, by a single nontrivial $\mathbb{Z}_2$ anyon. This \textit{intrinsically fermionic anomaly} permits anyon spins forbidden in bTOs: the simplest new member is an Abelian fTO with a single $\mathbb{Z}_2$ anyon $v$ of spin $1/8$ obeying $v^2=f$, with $f$ the \textit{local} fermion. Its anomaly lies in the $\nu=2$ layer of the $\mathbb{Z}_{16}$ family, and it has two ground states on the torus. While the generator is an Abelian anyon in the even layers, in the odd layers (those with a nontrivial Majorana) it instead has a non-Abelian fusion rule due to a Majorana zero mode, obeying $\tau^2 = 1\oplus f$. Since the fermion $f$ is local, the non-Abelian fusion rule still implies that $\tau$ generates an invertible $\Z_2$ 1-form symmetry~\footnote{This is reminiscent of the chiral $\Z_2$ symmetry in a non-chiral Majorana fermion in (1+1)D; its symmetry defect carries a Majorana zero mode and has the fusion rule $\tau\otimes \tau = 1
\oplus f$. Since the fermion $f$ is local, the corresponding symmetry operator $\mathcal{D}_\tau/\sqrt{2}$ up to normalization generates an invertible $\Z_2$ symmetry in the anti-periodic sector. Therefore this fusion rule is usually understood as an anomalous $\Z_2$ symmetry rather than a non-invertible symmetry. See e.g., \cite{inamurafermionic}.
It is analogous to our fusion rule of anyons with odd $\nu$ in the sixteen-fold way, where the anyon generates an anomalous $\Z_2$ symmetry with an anomaly in $\nu\in \Z_{16}$. }.

We show that each of these theories arises as the gapped boundary of a (3+1)D fermionic SPT protected by the $\mathbb{Z}_2$ 1-form symmetry. These bulk phases acquire their $\mathbb{Z}_{16}$ classification once the spacetime spin structure is \textit{twisted} by the 1-form symmetry.
We construct microscopic realizations as Walker--Wang lattice models coupled to local fermions, with the fTOs appearing on their boundaries.

\textit{New Abelian fermionic topological order in (2+1)D.} 
To construct a new instance of fTO in (2+1)D, we first consider a bosonic $U(1)_4$ Chern-Simons theory and couple it to local fermions in a specific manner. $U(1)_4$ has Abelian anyons $\{1,v,\psi, v\psi\}$ that generate a $\Z_4$ 1-form symmetry; $v$ carries spin 1/8, and $\psi$ is an emergent fermion with spin 1/2.
One route to obtaining a fermionic phase is by introducing a local fermion and then condensing the emergent fermion of $U(1)_4$; this procedure corresponds to gauging the $\Z_2$ 1-form symmetry generated by the emergent fermion $\psi$. The $\Z_2$ 1-form symmetry is anomalous but since its anomaly is trivialized by the spin structure of the fermionic theory, it can nonetheless be gauged. This follows the standard procedure of fermion condensation~\cite{Gaiotto:2015zta, bhardwajGaiottoKapustin2017, Aasen_2019, Tata2022anomalies}, and results in a fermionic invertible phase equivalent to the Chern insulator~\footnote{In this Letter, the Chern insulator denotes two copies of p+ip superconductors with $U(1)$ symmetry forgotten}.
However, to obtain the desired new topological order, we must perform the fermion condensation differently by coupling to a \textit{twisted} spin structure.

A spin structure of the spacetime is a trivialization of the second Stiefel-Whitney class $w_2$ of the spacetime, namely a choice of the $\Z_2$ cochain $\xi\in C^1(M,\Z_2)$ satisfying $d\xi=w_2$; this enables the Lorentz invariant theory to have local fermionic degrees of freedom. 
When the $\Z_2$ 1-form symmetry is present, writing the background gauge field $C\in Z^2(M,\Z_2)$, one can introduce a twisted spin structure by the trivialization $d\xi = w_2 + C$.

The background gauge field of $\Z_4$ 1-form symmetry in $U(1)_4$ can be represented by a pair of 2-form $\Z_2$ gauge fields $B,C$, where $dC=0$ and $dB=\frac{d\hat{C}}{2}$ mod 2, with $\hat C$ the $\Z_4$ lift of $C\in Z^2(M,\Z_2)$. $B$ corresponds to the $\psi$ lines, while $C$ represents $v$ lines.
We now couple $U(1)_4$ to twisted spin structure $d\xi=w_2+C$, and gauge the $\Z_2$ 1-form symmetry by summing over $B$:
\begin{equation}
    Z(C,\xi)=\sum_{B}z_\xi(B)Z_{U(1)_4} (B,C)~,
    \label{eq:gauged U(1)4}
\end{equation}
where $Z_{U(1)_4} (B,C)$ is the partition function of the $U(1)_4$ theory with $\Z_4$ backgrounds, and $z_\xi(B)$ is a certain phase factor regarded as a discrete torsion depending on twisted spin structure $\xi$. 
The 't Hooft anomaly of the $U(1)_4$ theory involving $B$ is given by the response $\pi \int B\cup C + Sq^2B$, reflecting the mutual braiding between $\psi, v$ and the self-statistics of $\psi$. Here we omitted the pure $C$ anomaly with the form $\frac{2\pi}{8}\int C\cup C + ...$ . 
Noting that $Sq^2 B = w_2\cup B$ on closed oriented manifolds, one can trivialize the $B$ anomaly $B\cup C + w_2\cup B = d(\xi\cup B)$ by twisted spin structure; the discrete torsion $z_\xi$ is a local counterterm that cancels this $B$ anomaly. 

After gauging the $\Z_2$ 1-form symmetry $B$, the resulting theory $Z(C,\xi)$ depends on twisted spin structure $\xi$ and has $\Z_2$ 1-form symmetry for $C$. This $\Z_2$ symmetry has a \textit{mod 8} 't Hooft anomaly $\frac{2\pi}{8}\int C\cup C + ...$ reflecting the spin 1/8 of $v$. This is illegally quantized as a bosonic 't Hooft anomaly; in bosonic phases the $\Z_2$ 1-form symmetry can only carry spin $n/4$ with $n\in\Z_4$. This implies that the $\Z_2$ 1-form symmetry of the resulting theory has an intrinsically \textit{fermionic} 't Hooft anomaly. Another way to see the symptom of the anomaly is that the $\Z_2$ anyons $v$ in the resulting theory fuse into a local fermion $f$: $v^2=f$, meaning that background gauge transformations of the 1-form symmetry violate fermion parity.

Since the fermions are local instead of emergent, the fusion rule $v^2=f$ still implies that $v$ generates $\Z_2$ 1-form symmetry. Because of the anomalous $\Z_2$ symmetry, the resulting theory $Z(C,\xi)$ cannot be invertible; this is in contrast to the standard fermion condensation with (untwisted) spin structure where we get the Chern insulator. In our case, the resulting theory is a nontrivial Abelian topological order with two states on the 2-torus.

\paragraph*{Mod 16 anomaly of $\Z_2$ 1-form symmetry in (2+1)D.}
The above 't Hooft anomaly of the $\Z_2$ 1-form symmetry in (2+1)D is classified by the reduced bordism group of 4D manifolds equipped with twisted spin structure, which is given by $\Z_{16}$. 
By anomaly inflow, the 't Hooft anomaly corresponds to the fermionic SPT phases with $\Z_2$ 1-form symmetry in (3+1)D which are classified through three layers of group cohomologies~\cite{WangGu2020, thorngren2019anomaliesbosonization, barkeshli2021invertible, Delmastro2021global}:
\begin{itemize}
    \item $H^4(B^2\Z_2, U(1))=\Z_4$ \quad (bosonic layer),
    \item $H^3(B^2\Z_2,\Z_2) = \Z_2$ \quad (complex fermion layer),
    \item $H^2(B^2\Z_2,\Z_2) = \Z_2$ \quad (Kitaev's Majorana layer).
\end{itemize}
The complete $\Z_{16}$ classification stems from the Majorana layer $\Z_2$ extended by the complex fermion layer $\Z_2$, further extended by the bosonic anomaly $\Z_4$. The anomalous theory obtained in Eq.~\eqref{eq:gauged U(1)4} carries the $\nu=2$ anomaly in $\Z_{16}$, lying in the complex fermion layer; indeed, fusing a pair of anyons creates a fermion parity odd state. 

Let us describe the (3+1)D topological response for the $\nu=2$ theory that describes the inflow. It is given by
\begin{align}
    Z^{\nu=2}_{\text{bulk}}(C,\xi)=z_\xi\left(\frac{d\hat C}{2}\right)\exp\left(\frac{2\pi i}{8} \int_M \hat C\cup \hat C + \hat C\cup_1d\hat C\right)
    \label{eq:nu=2CYtheory}
    \end{align}
where $\hat C$ is a $\Z_8$ lift of the $\Z_2$ gauge field $C$, and $z_\xi(\frac{d\hat C}{2})$ is a phase factor involving twisted spin structure $\xi$. This theory is explicitly defined through the local path integral of the Grassmann variables in the spacetime. This Grassmann integral has been developed for local definitions of fermionic SPT phases~\cite{gu2014, Gaiotto:2015zta}. See Supplemental Material (SM)~\cite{supmat} for details. The above theory $Z^{\nu=2}_{\text{bulk}}(C,\xi)$ describes the gauge invariant response of the (3+1)D invertible phase. The second term $\frac{2\pi}{8}\int C\cup C + ...$ accounts for spin 1/8 of the anyon. 

We now turn to the (3+1)D topological response of the $\nu=1$ theory. When a 4-manifold is equipped with the twisted spin structure $d\xi = w_2+C$, the 2-form $\Z_2$ gauge field $C\in Z^2(M,\Z_2)$ always admits a lift to an integral cocycle $\hat{C}\in Z^2(M,\Z).$ Using such an integral cocycle $\hat{C}$, the topological response for the $\nu=1$ theory is written as
\begin{align}\label{eq:nu1bulk}
    Z^{\nu=1}_{\text{bulk}}(C,\xi) = \exp\left(\frac{2\pi i}{16}\int \hat{C}\cup \hat{C}\right)\times (-1)^{\text{Arf}(F_{\hat{C}})}~.
\end{align}
Here, the 2D dual surface to $\hat{C}$ is denoted $F_{\hat{C}}$. The data of a 4-manifold $M$ along with a submanifold Poincaré dual to a class $\hat{C}$ on $M$ is an example of manifold structure dubbed a \textit{characteristic structure}, 
and we explain its precise definition and relevant properties in the SM. In particular, $F_{\hat{C}}$ is a spin manifold and serves as the obstruction from extending the spin structure on $M\backslash F$ to $M$.  We denote the mod 2 Arf invariant of $F_{\hat{C}}$ by $\text{Arf}(F_{\hat{C}})$. The factor $(-1)^{\text{Arf}(F_{\hat{C}})}$ is regarded as decorating Kitaev's Majorana chain on the dual surface, with the Arf invariant a bordism invariant that evaluates its partition function on $F_{\hat{C}}$.
This reflects that the $\nu=1$ theory is within Kitaev's Majorana layer.

The partition function $Z^{\nu=1}_{\text{bulk}}(C,\xi)$ gives the signature of the 4-manifold mod 16: $Z^{\nu=1}_{\text{bulk}}(C,\xi)=\exp(\frac{2\pi i}{16}\sigma(M))$ and defines a gauge invariant response~\cite{guillou, KirbyTaylor}. Note that, while the $\nu=1$ theory is independent of the twisted spin structure $\xi$, it nevertheless requires a twisted spin structure to be well-defined and hence fundamentally remains a twisted spin theory. Using the lift to integral cocycle $\hat{C}\in Z^2(M,\Z)$, the $\nu=2$ theory in \eqref{eq:nu=2CYtheory} is simply written as 
\begin{align}
    Z_{\nu=2}(C,\xi) = \exp\left(\frac{2\pi i}{8}\int \hat{C}\cup \hat{C}\right) = (Z_{\nu=1})^2~,
\end{align}
therefore the two copies of the $\nu=1$ SPT phases indeed produce the $\nu=2$ phase.

\begin{figure}[htb]
    \centering
    \includegraphics[width=0.9\linewidth]{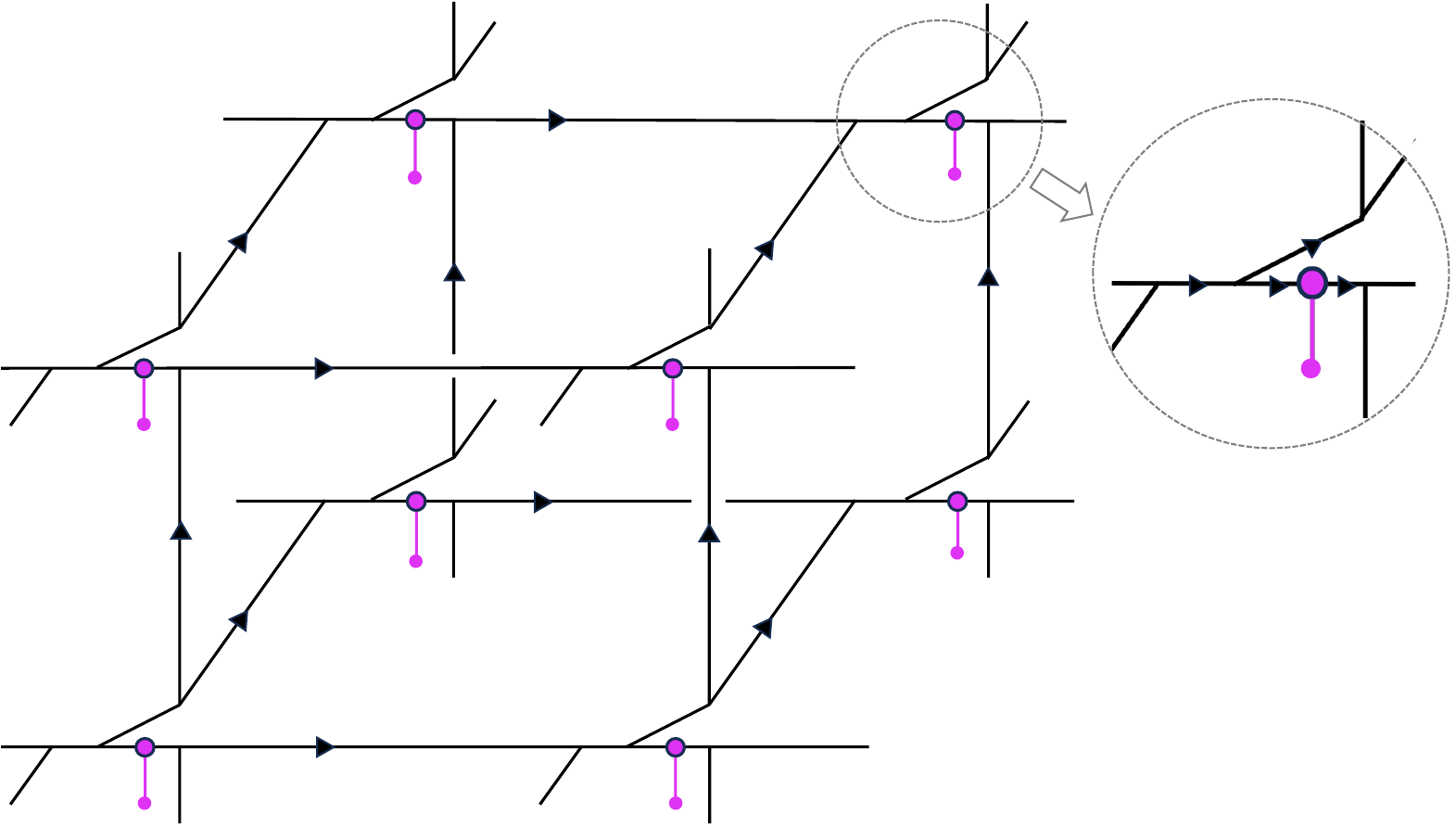}
    \caption{The cubic lattice for defining the wave function. The long edges (directed black edges in the main figure) are labeled by anyons in $\{1,v\}$, while the short edges (four directed black edges expanded on the right) are labeled by $\{1,v,\psi,v\psi\}$. The black diagram is not closed under fusion by itself, and could fuse into a pink line labeled by $\{1,\psi\}$. To obtain a closed diagram, the pink lines are connected arbitrarily \change{along cubic lattice edges} 
    (see Fig.~\ref{fig:grassmann_cube}).}
    \label{fig:walkerwang}
\end{figure}

\paragraph*{Walker-Wang type lattice model for $\nu=2$ theory.}
We provide a Hamiltonian model for the $\nu=2$ theory described in Eq.~\eqref{eq:gauged U(1)4}. The Hamiltonian is defined on a cubic lattice with vertices resolved into trivalent junctions (see Fig.~\ref{fig:walkerwang}). This setup is similar to the original Walker-Wang construction~\cite{walkerwang2013}, but additionally coupled to local complex fermions. Let us describe the degrees of freedom:
\begin{itemize}
    \item Among the black edges of the resolved cubic lattice in Fig.~\ref{fig:walkerwang}, we call the edges of the original cubic lattice before resolution the ``long edges'', and the edges appearing after resolutions ``short edges''. We put a single qubit on each long edge of the lattice, spanned by states $\{\ket{1}, \ket{v}\}$.
    \item We put a $\Z_4$ qudit on each short edge, spanned by the states $\{\ket{1}, \ket{v}, \ket{\psi},\ket{v\psi}\}$.
    \item We put a complex fermion on each pink dot introduced at each vertex of the cubic lattice.
    \end{itemize}

\begin{figure}[htb]
    \centering
    \includegraphics[width=0.9\linewidth]{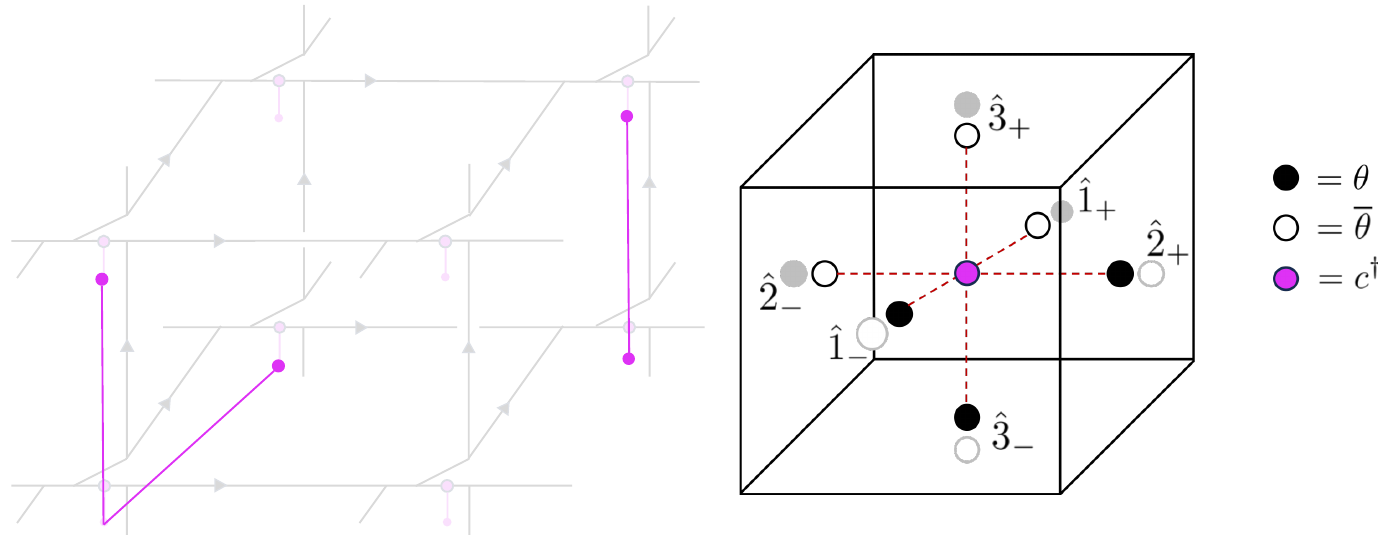}
    \caption{Left: Given that the assignment of $\{1,v\}$ on long edges is identified with $C\in Z^2(\hat{\mathcal{T}},\Z_2)$, pink $\psi$ lines are stemming from the diagram according to the 3-form $d\hat{C}/2$. The $\psi$ lines are then connected along cubic lattice edges according to its trivialization; $B\in C^2(\hat{\mathcal{T}},\Z_2 )$ satisfying $dB = d\hat{C}/2$. Right: assignment of Grassmann variables on each face of the dual cubic lattice.}
    \label{fig:grassmann_cube}
\end{figure}

Similarly to the Walker-Wang model, the ground-state wave function is a superposition of admissible anyon diagrams of the $U(1)_4$ theory $\{1,v,\psi,v\psi\}$ on the resolved cubic lattice. This connects anyons on the edges labeled as follows:
\begin{itemize}
    \item We put anyons in $\{1,v\}$ on black long edges of the resolved cubic lattice. We require that on each vertex of the cubic lattice, the number of long edges labeled by $v$ gathering at the vertex is even. Due to the Poincar\'e duality, the assignment of such $\{1,v\}$ configurations is identified as an element $C\in Z^2(\hat{\mathcal{T}},\Z_2)$, where $\hat{\mathcal{T}}$ is a dual lattice of the cubic lattice.
    \item We put anyons in $\{1,v,\psi,v\psi\}$ on black short edges of the resolved cubic lattice. Once we impose the admissible fusion rule on each trivalent junction, the anyon labels on short edges are uniquely determined from those on long edges.
    \item On each pink dot, we allow two short edges to fuse into a pink line valued in $\{1,\psi\}$: the pink line moves \change{along the cubic lattice edge, and connected to the top} 
    (see Fig.~\ref{fig:grassmann_cube}). 
    The vertex with $\psi$ line is identified as $\frac{d\hat C}{2}\in Z^3(\hat{\mathcal{T}},\Z_2)$, where $\hat{C}$ denotes the $\Z_4$ lift of $C\in Z^2(\hat{\mathcal{T}},\Z_2)$. The pink lines are not associated with states in the Hilbert spaces, hence are not dynamical degrees of freedom.
    \end{itemize}

Since  $\frac{d\hat C}{2}\in Z^3(\hat{\mathcal{T}},\Z_2)$ is trivial in cohomology on a cubic lattice, there is a trivialization $dB=\frac{d\hat C}{2}$ using $B\in C^2(\hat{\mathcal{T}},\Z_2)$. We identify this $B$ as the dual of pink lines in Fig.~\ref{fig:grassmann_cube}. After all, the anyon diagram is labeled by a pair of cochains $(B,C)$, and we write its evaluation as $Z_{\text{WW}}(B,C)$.

To complete the description of the wave function, we illustrate the wave function of complex fermions on each pink dot. The wave function is described through the Grassmann integral on a dual cubic lattice $\hat{\mathcal{T}}$, introduced in \cite{Chen2023highercup}. We introduce a pair of Grassmann variables $\{\theta, \overline\theta\}$ on each face of the dual cubic lattice $\hat{\mathcal{T}}$, as in Fig.~\ref{fig:grassmann_cube}. The Grassmann integral takes the form
\begin{align}
    \sigma(B,C) = \int\prod_f d\vartheta_f d \overline\vartheta_f\prod_{\Delta} u(\Delta)~,
\end{align}
where $u(\Delta)$ is the Boltzmann weight on each cube $\Delta$ of the dual lattice $\hat{\mathcal{T}}$ that contains a complex fermion operator $c_\Delta^\dagger$ on the pink dot:
\begin{align}
    u(\Delta) = \theta_{\hat 1_-}^{B(\hat 1_-)}\theta_{\hat 2_+}^{B(\hat 2_+)}\theta_{\hat 3_-}^{B(\hat 3_-)} 
    \overline\theta_{\hat 1_+}^{B(\hat 1_+)}\overline\theta_{\hat 2_-}^{B(\hat 2_-)}\overline\theta_{\hat 3_+}^{B(\hat 3_+)} (c_\Delta^\dagger)^{\frac{d\hat C}{2}(\Delta)}
\end{align}
Since the complex fermion operator $c^\dagger$ is Grassmann odd and $dB=\frac{d\hat C}{2}$, $u(\Delta)$ is Grassmann even. After integrating the Grassmann variables $\theta,\overline\theta$, $\sigma(B,C)$ becomes a product of fermion operator $\prod_\Delta(c_\Delta^\dagger)^{\frac{d\hat C}{2}(\Delta)}$ up to a $(\pm 1)$ phase. \change{The Grassmann variables $(\vartheta_f, \overline{\vartheta}_f)$ become $(\theta_f, \overline{\theta}_f)$ or $(\overline{\theta}_f, \theta_f)$, depending on the configuration of $C$; see SM \cite{supmat} for details.}

Then, the ground state wave function is given by
\begin{align}
    \ket{\Psi} = \sum_{C\in Z^2(\hat{\mathcal{T}},\Z_2)}Z_{\text{WW}}(B,C)\ket{C}_{\text{edges}} \otimes \sigma(B,C)\ket{0}_{\text{F}}~,
\end{align}
where $\ket{C}_{\text{edges}}$ denotes the state of the bosonic Hilbert space formed by qubits and qudits on edges. Since $C\in Z^2(\hat{\mathcal{T}},\Z_2)$ uniquely determines the admissible anyon diagram, it labels the state with the corresponding anyon diagram $\ket{C}$. $\ket{0}_{\text{F}}$ is a fermionic Fock vacuum of the complex fermions on pink dots, on which the fermions are created by the operator $\sigma(B,C)$.

Importantly, one does \textit{not} have to sum over $B$ in the above expression; for each given configuration of $C$, one can make any choice of $B$ satisfying $dB=\frac{d\hat{C}}{2}$ and use it in the expression, and then the state $\ket{\Psi}$ is independent of the choice of $B.$ This is crucial since $B$ fields are not associated with dynamical degrees of freedom. The Hamiltonian is then given by
\begin{align}
    H_{\nu=2} = -\sum_v \mathcal{P}_v - \sum_p \mathcal{P}_p~,
\end{align}
where $\mathcal{P}_v$ is a projector enforcing the $U(1)_4$ fusion rule of each fusion vertex.
$\mathcal{P}_p$ is defined on each plaquette of the resolved cubic lattice; it shifts the anyon diagram by a small loop of $v$ along $\partial p$, and acts on complex fermion operators accordingly (see SM~\cite{supmat} for details).

When this model is defined on a manifold with boundary, the boundary has a deconfined excitation created by the $v$ line, and it behaves as a $\Z_2$ anyon with the fusion rule $v\times v = f$, with $f$ a local fermion. Therefore, the surface topological order realizes the desired fTO described in Eq.~\eqref{eq:gauged U(1)4}.

\paragraph*{Sixteen-fold way of fTOs in (2+1)D and their anomalies.}
We now give an explicit construction for the topological order that saturates the $\nu=1$ anomaly in $\Z_{16}$. By the theory of minimal nondegenerate extensions (see SM~\cite{supmat}), the non-spin theory corresponding to $\nu=1$ is given by $Spin(1)_1$, with anyons $\{1,\psi,\sigma\}$. The construction leading to the fTO with $\nu=2$ anomaly cannot be applied directly in the present setting, as the anyon with spin 1/16 is non-Abelian. Nonetheless, there exists a different mechanism for coupling the theory to a (twisted) spin structure. This construction becomes illuminating upon stacking two copies of the theory, which reveals a connection to the $\nu=2$ phase.

Consider the algebra in $Spin(1)_1$ given by $\mathcal A=(1\oplus \psi)$, which is only commutative up to a minus sign. We compute the category of  modules of $Spin(1)_1$ with respect to this algebra as follows: 
\begin{equation}
    \cA \otimes 1 = \cA, \quad \cA \otimes \psi = \cA, \quad \cA \otimes \sigma = \sigma \oplus \psi \sigma\,.
\end{equation}
Passing to the category of local modules, this defines a new theory $\mathcal{T}$. In this theory, the algebra $\cA$ is equivalent to the vacuum $\mathds{1}$ and local fermion $f$; thus, the theory is genuinely fermionic. Further, the sum of simple objects $(\sigma \oplus f\sigma)$ becomes a new simple object which we denote by $\tau$, with the property that $\tau$ braids with the local fermion with a minus sign and that $\tau\otimes \tau = \mathds{1}$\, \footnote{Unlike in the case of boson condensation, the objects in an orbit for a fermionic condensation differ by spin $1/2$ \cite{Yu:2021zmu}.}. Physically, $\tau$ is a vortex of fermion parity carrying a Majorana zero mode, reflecting that the $\nu=1$ theory has the anomaly within the Majorana layer. Fusing a pair of $\tau$ produces a sum of the vacuum and local fermion $f$, hence $\tau$ generates an anomalous $\Z_2$ 1-form symmetry in this theory. 

It is meaningful to consider inserting $\tau$ along a Poincaré dual to $w_2(TM)$. With $\tau$ insertion, braiding between a local fermion $f$ and the Poincar\'e dual of $w_2(TM)$ becomes trivial; the $(-1)$ phase from the framing anomaly carried by $f$ and $(-1)$ mutual braiding between $f$ and $\tau$ cancels out. This implies that $\tau$ in the theory becomes an anyon in the untwisted sector of spin TQFT (which in fact couples to \textit{twisted} spin structure, as we will see shortly).

The insertion of $\tau$ along a Poincaré dual to $w_2(TM)$ implies that the theory couples to a twisted spin structure, where the twist is by the background field for the the $\Z_2$ 1-form symmetry. Note that the ambient manifold $M$ in which the (2+1)D theory is defined is oriented, since the theory $Spin(1)_1$ that we began with is bosonic. The data of a manifold $M$ along with a submanifold $F$ that is Poincaré dual to $w_2(TM)$ is known as a {Freedman-Kirby characteristic structure} \cite{FK} and, as discussed in the SM~\cite{supmat}, such a characteristic structure is equivalent to a $spin^c$ structure. Writing this as a twisted spin structure given by $w_2(TM) = c_1(L) \mod 2$, we can now map the class $c_1$ into $H^2( B^2\Z_2,\Z_2)$ by mod 2 reduction, where it serves as the background for the $\Z_2$ 1-form symmetry. 
In summary, the theory $\mathcal T$ has an anyon $\tau$ carrying the fermion parity vortex generating a $\Z_2$ 1-form symmetry, and a twisted spin structure after inserting $\tau$ along a Poincaré dual to $w_2(TM)$.

Now consider stacking two copies of $\mathcal{T}$: since they each carry the anomaly $\nu=1$, the resulting stacked theory should be equivalent, up to a gapped interface, to the theory with $\nu=2$ anomaly given in Eq.~\eqref{eq:gauged U(1)4}. Since the theory $\mathcal{T} \boxtimes \mathcal{T}$ is a tensor of two fermionic theories, it is still fermionic i.e., contains a local fermion in the spectrum. Its other objects are given by 
\begin{align}
    \text{ob}\left({\cT \boxtimes \cT}\right)=  &\{ f \boxtimes f, \quad  \tau \boxtimes \tau,\quad
    \tau \boxtimes f , \\ \notag 
    &\,\,f \boxtimes \tau , \quad \tau \boxtimes \mathds{1}, \quad \mathds{1} \boxtimes \tau \}\,.
\end{align}
The object $f \boxtimes f$ takes the the bosonic vacuum from both theories denoted by $\mathds{1}$, and stacks on the local fermion $f$ within each theory. The object $\tau \boxtimes f$ is more explicitly given by $\tau$ stacked with the vacuum of the second theory $\mathcal T$, in which the vacuum is dressed by the local fermion. $\mathcal T \boxtimes \mathcal T$ also has the operator $\mathds{1} \boxtimes \mathds{1}$ from taking  $f \boxtimes f$  and forgetting the local fermions. This allows us to  consider the algebra $\mathcal B =(\mathds{1}\boxtimes \mathds{1} \oplus f \boxtimes f)$, which we can condense like a boson in this composite theory. Upon condensing this boson, the objects $(\tau \boxtimes f \oplus \tau \boxtimes \mathds{1})$ and $(f\boxtimes \tau \oplus \mathds{1}\boxtimes \tau)$ are projected out, and we are left with two copies of $\tau \boxtimes \tau$. We then identify $\mathcal{B}$ with the vacuum for the theory after condensation, which we denote $\mathbf{1}$, via the maps $(\mathds{1}\boxtimes \mathds{1} \oplus f \boxtimes f) \to \mathbf{1}\oplus \mathbf{1}\to \mathbf{1}\,.$

We also identify the local fermions from both theories into a single local fermion, still denoted by $f$. In a fully bosonic theory, upon making the identification of the algebra $\mathcal B$ with the vacuum, the two copies of $\tau \boxtimes \tau$ become two distinguished simple objects. However, because we are still in an ambient fermionic setting  with a local fermion (which cannot be changed on condensing a boson), we can create a new object by considering the composite $(\tau \boxtimes \tau) \oplus f(\tau \boxtimes \tau)$, where the two objects in the sum differ by spin $1/2$.
This is because, after condensing the boson, in the fermionic theory we also have a map $\mathbf{1}\to \mathbf{1} \oplus \mathbf{1}\to \mathbf{1}\oplus f$
that involves stacking by the local fermion. As a module for $\mathbf{1}\oplus f$, we define $(\tau \boxtimes \tau) \oplus f(\tau \boxtimes \tau)$ to be an anyon $\alpha$ which is a simple object in this category. In particular, in a theory with a purely bosonic vacuum $\alpha$ decomposes into two irreducibles, but when we allow the vacuum to be stacked by a local fermion, $\alpha$ itself remains irreducible.

In conclusion, the category after condensing $\mathcal B$ has the following objects
\begin{equation}
   \text{ob}\left(\frac{\cT \boxtimes \cT}{\mathcal B}\right) = \{\mathbf{1},f\} \times \{ \alpha\}
\end{equation}
with the following properties: 
\begin{itemize}
    \item There is a local fermion in the spectrum, forcing the theory to couple to spin structure. The anyon $\alpha$
   implements a $\Z_2$ 1-form symmetry.
   \item Using the composite $\tau \boxtimes \tau$, the anyon $\alpha$  has spin $1/8$.
    \item By wrapping $\alpha$ on a Poincaré dual to $w_2(TM)$, the theory $\mathcal{T}\boxtimes \mathcal{T}/\mathcal{B}$ actually couples to a twisted spin structure for the $\Z_2$ 1-form symmetry, in the same way that $\mathcal T$ does.    
\end{itemize}
This description is equivalent to the theory described by $Z(C,\xi)$ in Eq.~\eqref{eq:gauged U(1)4}.

The following commuting square summarizes the relevant manipulations on the (2+1)D boundary theories:
\begin{equation}\label{eq:bulkboundary}
    \begin{tikzcd}
    \mathcal{T} \arrow[dd,"\frac{- \boxtimes \mathcal T}{\mathcal{B}}"]& &\arrow[ll,"\text{Condense $\cA$}",swap] Spin(1)_1\arrow[dd,"\frac{- \boxtimes Spin(1)_1}{\mathcal{D}}"]\\
    && \\
    \frac{\mathcal{T}\boxtimes \mathcal{T}}{\mathcal{B}}& & \arrow[ll,"\text{Fermionize}"] Spin(2)_1
\end{tikzcd}
\end{equation}
In particular, starting with the $Spin(1)_1$ theory, we form a fermionic theory $\cT$ that saturates the $\nu=1$ anomaly. Then, the fermionic theory for the $\nu=2$ anomaly can be obtained in two equivalent ways. The first, indicated by the left downward arrow, is simply to stack by $\cT$ and condense a bosonic algebra $\mathcal{B}$. Alternatively, the same theory is obtained via the downward right arrow, starting from $Spin(1)_1$ and fermionizing using $z_\xi(B)$ in Eq.~\eqref{eq:gauged U(1)4}. In particular, the algebra $\mathcal{D}$  we condense is bosonic and given by $(\mathds{1}\boxtimes \mathds{1} \oplus \psi \boxtimes \psi)$, arising from the composite of the two emergent fermions carried by both theories \cite{Teixeira:2025qsg}. The right-hand side of the diagram reflects how the usual sixteen-fold way is recovered i.e., by successively applying the same operation. The central theme of this paper is the passage to the left-hand side, where the analogous construction is carried out in a purely fermionic setting.

\paragraph*{Discussion.} Our work opens several future directions. First, it would be desirable to develop lattice Hamiltonian realizations for the whole $\Z_{16}$ family. Here, we presented a Walker-Wang type construction for the $\nu=2$ phase, whose boundary realizes the Abelian fTO with a $\Z_2$ anyon of spin $1/8$. The generator $\nu=1$ is more involved, since it lies in the Majorana layer of the classification. The bulk SPT phase would involve decoration by Kitaev's Majorana chains. 

While we have provided a new sixteen-fold family of fTOs, we can generate another family of topological orders by coupling bTOs with the fermionic sixteen-fold way, and then condensing the diagonal anyons. It would be interesting to systematically develop the classification of topological orders coupled to twisted spin structure. The phase transitions between distinct fTOs also remain to be understood.

Especially important is understanding whether the fermionic anomalies presented here can arise in strictly two-dimensional lattice models without coupling to any bulk. If such 2D models exist, our results call for a direct lattice diagnostic of fermionic higher-form anomalies, akin to how bosonic 't Hooft anomalies on the lattice can be determined for a given symmetry operator~\cite{Else:2014vma, Feng2026higher}. Formulating a general anomaly index that extracts our mod $16$ anomaly presents an open challenge. 

\section*{Acknowledgments}
We thank Shu-Heng Shao for discussions.
RK is supported by the Department of Applied Physics,
the University of Tokyo. 
AP is supported by the U.S. Department of Energy, Office of Science, Office of Advanced Scientific Computing Research, through the Exploratory Research for Extreme Scale Science (EXPRESS) program under Award No. DE-SC0026216.
MY is supported by the EPSRC Open Fellowship EP/X01276X/1.
We thank the Galileo Galilei Institute for Theoretical Physics for
hosting the program “Defects and Extended Excitations in Quantum Field Theory, Quantum Matter and Statistical Models” in 2026, during which part of the work was completed.

\bibliography{bibliography.bib}

\onecolumngrid

\vspace{0.3cm}

\newpage

\onecolumngrid 
\clearpage
\makeatletter 

\begin{center}   
	\textbf{\large Supplementary Material for ``Sixteen-Fold Way for Fermionic Topological Orders"}\\
	[1em]
	Ryohei Kobayashi$^1$, Abhinav Prem$^{2}$, and Matthew Yu$^{3}$, \\[.1cm]
	{\itshape \small ${}^1$Department of Physics, The University of Tokyo, 7-3-1 Hongo, Bunkyo-ku, Tokyo, 113-0033, Japan\\ 
	${}^2$Physics Program, Bard College, 30 Campus Road, Annandale-on-Hudson, New York 12504, USA\\
    ${}^4$Mathematical Institute, University of Oxford, OX2 6GG, Oxford, UK}\\
	(Dated: \today)\\[1cm]
\thispagestyle{titlepage} 
\end{center} 	

\renewcommand{\thefigure}{S\arabic{figure}}
\setcounter{figure}{0}

\appendix
\setcounter{secnumdepth}{2} 


This Supplemental Material contains a number of Appendices with technical details supporting the results presented in the main text. 


\section{Detailed description of the $\nu=2$ theory}

\subsection{Review of the Grassmann integral}
\label{subsec:grassmann}

We review the Grassmann integral $\sigma(M^d,B_{d-1})$, defined in terms of Grassmann variables on a triangulated manifold. This construction follows the path-integral formulation of Grassmann variables on triangulations developed in Refs.~\cite{gu2014,Gaiotto:2015zta,Kobayashi2019pin,Tata2022anomalies}. Throughout this discussion, $M^d$ is a closed triangulated $d$-manifold equipped with a branching structure. The branching induces an orientation sign $\epsilon(\Delta_d)=\pm 1$ for each $d$-simplex $\Delta_d$. The Grassmann integral depends on a $(d-1)$-cocycle
\begin{align}
B_{d-1} &\in Z^{d-1}(M^d,\Z_2).
\end{align}
For simplicity, we restrict our attention to the case where $M^d$ is oriented.

Given a $(d-1)$-form background $B_{d-1}$, we introduce Grassmann variables only on those $(d-1)$-simplices $\Delta_{d-1}$ for which
\begin{align}
B_{d-1}(\Delta_{d-1}) &= 1.
\end{align}
More precisely, for every such simplex we assign a pair of Grassmann variables
\begin{align}
\theta_{\Delta_{d-1}},
\qquad
\overline{\theta}_{\Delta_{d-1}}.
\end{align}
These two variables are placed on the two sides of $\Delta_{d-1}$, namely in the two neighboring $d$-simplices sharing that face. The precise convention determining which side carries $\theta$ and which carries $\overline{\theta}$ will be specified below.

The Grassmann integral is then defined by
\begin{align}
\sigma(M^d,B_{d-1})
&=
\int
\prod_{\Delta_{d-1}: B_{d-1}(\Delta_{d-1})=1}
d\theta_{\Delta_{d-1}},d\overline{\theta}_{\Delta_{d-1}}
\prod_{\Delta_d} u(\Delta_d),
\label{eq:sigma_def}
\end{align}
where $u(\Delta_d)$ denotes the ordered product of the Grassmann variables associated with the $(d-1)$-faces of $\Delta_d$. The product over $\Delta_d$ runs over all $d$-simplices of the triangulation.

As a simple example, consider $d=2$ and a 2-simplex $\Delta_2=\langle 012\rangle$. Then $u(\Delta_2)$ is a product of the Grassmann variables associated with the three edges $\langle 12\rangle$, $\langle 01\rangle$, and $\langle 02\rangle$, each appearing only when the corresponding value of $B$ is nonzero. Thus, for a positively oriented simplex, one has
\begin{align}
u(012)
&=
\vartheta_{12}^{B(12)}
\vartheta_{01}^{B(01)}
\vartheta_{02}^{B(02)},
\end{align}
where each $\vartheta$ denotes either $\theta$ or $\overline{\theta}$, depending on the assignment convention described below. Since $B_{d-1}$ is closed, the number of Grassmann variables appearing in each $u(\Delta_d)$ is even. Hence $u(\Delta_d)$ is Grassmann-even.

The fermionic signs arising from reordering Grassmann variables imply that $\sigma(M^d,B_{d-1})$ behaves as a quadratic function of $B_{d-1}$. The corresponding quadratic refinement depends on the ordering convention used in defining $u(\Delta_d)$. We use the ordering convention of Gaiotto and Kapustin~\cite{Gaiotto:2015zta}, which we now describe.

Let
\begin{align}
\Delta_d &= \langle 0,1,\cdots, d\rangle
\end{align}
be a $d$-simplex with vertices ordered by the branching structure. We denote by
\begin{align}
\hat{i} &:= \langle 0,1,\cdots,\widehat{i},\cdots, d\rangle
\end{align}
the $(d-1)$-face obtained by omitting the vertex $i$. For a positively oriented simplex, $\epsilon(\Delta_d)=+$, the Grassmann variables in $u(\Delta_d)$ are ordered by first listing the even faces in increasing order and then the odd faces in increasing order:
\begin{align}
\hat{0}
&\to \hat{2}\to \hat{4}\to \cdots
\to
\hat{1}\to \hat{3}\to \hat{5}\to \cdots .
\end{align}
For a negatively oriented simplex, $\epsilon(\Delta_d)=-$, the reverse ordering is used:
\begin{align}
\cdots
&\to \hat{5}\to \hat{3}\to \hat{1}
\to
\cdots \to \hat{4}\to \hat{2}\to \hat{0}.
\end{align}

For example, when $d=2$ and $\epsilon(\langle 012\rangle)=+$, this prescription gives
\begin{align}
u(012)
&=
\vartheta_{12}^{B(12)}
\vartheta_{01}^{B(01)}
\vartheta_{02}^{B(02)}.
\end{align}
On the other hand, for $\epsilon(\langle 012\rangle)=-$, the order is reversed:
\begin{align}
u(012)
&=
\vartheta_{02}^{B(02)}
\vartheta_{01}^{B(01)}
\vartheta_{12}^{B(12)}.
\end{align}

\begin{figure}[htb]
\centering
\IfFileExists{grassmann.pdf}{
\includegraphics[width=0.5\linewidth]{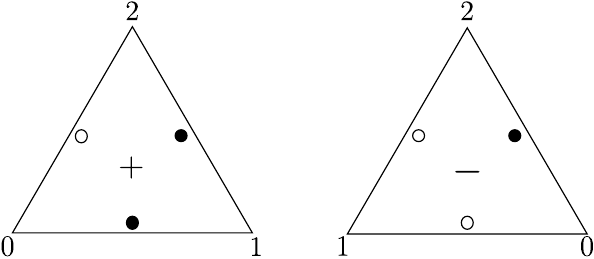}
}{
\fbox{\parbox{0.45\linewidth}{\centering Placeholder for \texttt{grassmann.pdf}}}
}
\caption{Convention for assigning Grassmann variables to 1-simplices in two dimensions. Black dots denote $\theta$, while white dots denote $\overline{\theta}$.}
\label{fig:Grassmann}
\end{figure}

It remains to specify whether $\theta_{\hat{i}}$ or $\overline{\theta}_{\hat{i}}$ appears for a given face $\hat{i}$. Our convention is as follows. For a positively oriented $d$-simplex $\Delta_d=\langle 0,1,\cdots,d\rangle$, the factor associated with $\hat{i}$ is $\overline{\theta}_{\hat{i}}$ when $i$ is odd, and $\theta_{\hat{i}}$ when $i$ is even. For a negatively oriented simplex, the assignment is reversed: $\theta_{\hat{i}}$ appears for odd $i$, while $\overline{\theta}_{\hat{i}}$ appears for even $i$. Equivalently, for $\epsilon(\Delta_d)=+$,
\begin{align}
u(\Delta_d)
&=
\theta_{\hat{0}}^{B(\hat{0})}
\theta_{\hat{2}}^{B(\hat{2})}
\theta_{\hat{4}}^{B(\hat{4})}
\cdots
\overline{\theta}_{\hat{1}}^{B(\hat{1})}
\overline{\theta}_{\hat{3}}^{B(\hat{3})}
\overline{\theta}_{\hat{5}}^{B(\hat{5})}
\cdots .
\end{align}
For $\epsilon(\Delta_d)=-$, we instead have
\begin{align}
u(\Delta_d)
&=
\cdots
\theta_{\hat{5}}^{B(\hat{5})}
\theta_{\hat{3}}^{B(\hat{3})}
\theta_{\hat{1}}^{B(\hat{1})}
\cdots
\overline{\theta}_{\hat{4}}^{B(\hat{4})}
\overline{\theta}_{\hat{2}}^{B(\hat{2})}
\overline{\theta}_{\hat{0}}^{B(\hat{0})}.
\end{align}
For $d=2$, this assignment may be represented pictorially as in Fig.~\ref{fig:Grassmann}, where black and white dots denote $\theta$ and $\overline{\theta}$, respectively.

Finally, the definition in Eq.~\eqref{eq:sigma_def} is independent of the ordering chosen for the $d$-simplices in the product. Indeed, since $B_{d-1}$ is a cocycle, $dB_{d-1}=0$, each local factor $u(\Delta_d)$ contains an even number of Grassmann variables and is therefore Grassmann-even. Thus the local factors commute with one another, making the global product well defined.

\subsection{Properties of the Grassmann integral}

The above Grassmann integral satisfies the following quadratic property:
\begin{equation}
    \sigma(B)\sigma(B')=\sigma(B+B')(-1)^{\int B\cup_{d-2}B'}~,
    \label{eq:ClosedQuad}
\end{equation}
which is directly derived by using anti-commutation relation of Grassmann variables on each Boltzmann weight $u(\Delta_d)$~\cite{Gaiotto:2015zta}.
This quadratic property implies that the Grassmann integral is not invariant under gauge transformations of the $(d-1)$-form $B$:
\begin{align}
    \sigma(B+d\chi) = \sigma(B)\cdot \sigma(d\chi) (-1)^{\int B\cup_{d-2}d\chi}~.
\end{align}
One can also see that
\begin{align}
    \sigma(d\chi) = (-1)^{\int \chi\cup_{d-4}\chi + \chi\cup_{d-3}d\chi + w_2\cup \chi}~,
    \label{eq:w2}
\end{align}
where $w_2$ is a specific representative of the second Stiefel-Whitney class of $M^d$ defined combinatorially from the branching structure. The above phase ambiguity under the gauge transformation corresponds to the inflow of the bulk $(d+1)$D response:
\begin{align}
    (-1)^{\int Sq^2B + w_2\cup B}~.
\end{align}

\subsection{Coupling to twisted spin structure}

One can couple the Grassmann integral to the twisted spin structure $d\xi = w_2 +C$, and define a theory that depends on the twisted spin structure:
\begin{align}
    z_\xi(B) = \sigma(B)\cdot (-1)^{\int\xi\cup B}~.
\end{align}
This theory has an 't Hooft anomaly given by the response
\begin{align}
    (-1)^{\int Sq^2B + B\cup C}~.
\end{align}

\subsection{$\nu=2$ theory}

Now we are ready to describe the path integral of the $\nu=2$ theory.
The (3+1)D topological response at $\nu=2$ is given by
\begin{align}
    Z_{\nu=2}(C,\xi)=z_\xi\left(\frac{d\hat{C}}{2}\right)\cdot \exp\left(\frac{2\pi i}{8} \int_M \hat C\cup \hat C + \hat C\cup_1d\hat C\right)
    \label{eq:nu=2CYtheory app}
    \end{align}
    where $\hat C$ is a $\Z_8$ lift of the $\Z_2$ gauge field $C$.

    The above theory gives a gauge invariant response; the 't Hooft anomaly of the term $z_\xi\left(\frac{d\hat{C}}{2}\right)$ is
    \begin{align}
        \pi\int Sq^2\left(\frac{d\hat{C}}{2}\right) +  C\cup \frac{d\hat{C}}{2}~,
    \end{align}
    This anomaly is canceled by the additional term $\frac{2\pi i}{8} \int_M \hat C\cup \hat C + \hat C\cup_1d\hat C$, since
    \begin{align}
       \frac{2\pi i}{8}   d\left( \hat C\cup \hat C + \hat C\cup_1d\hat C \right) = \pi \left(Sq^2\left(\frac{d\hat{C}}{2}\right) +   C\cup\frac{d\hat C}{2}\right)~.
    \end{align}
    Therefore the combination $Z_{\nu=2}(C,\xi)$ is gauge invariant as a whole.
The above theory is  a (3+1)D invertible theory with twisted spin structure $d\xi=w_2+C$, and describes the inflow of the fermionized $U(1)_4$ theory~\eqref{eq:gauged U(1)4} described in the main text.

We note that with the twisted spin structure, $d\hat C/2$ is always trivial in cohomology, therefore the theory in \eqref{eq:nu=2CYtheory app} does not depend on the choice of the twisted spin structure $\xi\to \xi+a$ with $a\in Z^1(M,\Z_2)$. Nevertheless it requires the twisted spin structure to be defined, and therefore is a twisted spin TQFT.

\section{Detailed description of the Walker-Wang type model}
In this Appendix, we provide additional details of the Hamiltonian model discussed in the main text.

\subsection{Detailed description of $\sigma(B,C)$}

\label{subsec:sigma(B,C)}

\change{Here we provide the detailed definition the operator $\sigma(B,C)$ introduced in the main text. 
We will define $\sigma(B,C)$ when $C$ is exact, $C=d\lambda$ using $\lambda\in C^1(\hat{\mathcal{T}},\Z_2)$. 
We do not attempt to define $C$ when $C$ is nontrivial in cohomology; we will define the wave function when $\hat{\mathcal{T}}$ is supported on a 3-sphere or a 3-ball, so that $C$ generally becomes exact.
Accordingly, we do not attempt to obtain a surface topological order on a nontrivial topology such as a 2-torus or demonstrate topological degeneracy. Rather, we demonstrate surface topological order by the presence of topological line operators and deconfined excitations at the boundary. }

\begin{itemize}
    \item First of all, when $C=0$, $\sigma(B,C=0)$ is an overall phase given by the Grassmann integral using $B\in Z^2(\hat{\mathcal{T}},\Z_2)$, 
\begin{align}
    \sigma(B,C=0) = \int\prod_f d\theta_f d \overline\theta_f\prod_{\Delta} u(\Delta)~,
\end{align}
with
\begin{align}
    u(\Delta) = \theta_{\hat 1_-}^{B(\hat 1_-)}\theta_{\hat 2_+}^{B(\hat 2_+)}\theta_{\hat 3_-}^{B(\hat 3_-)} 
    \overline\theta_{\hat 1_+}^{B(\hat 1_+)}\overline\theta_{\hat 2_-}^{B(\hat 2_-)}\overline\theta_{\hat 3_+}^{B(\hat 3_+)} 
\end{align}
which is Grassmann even due to $dB=d\hat C/2 = 0$. 
Using the expression $B=d\chi$ with $\chi\in C^1(\mathcal{T},\Z_2)$, the Grassmann integral satisfies \cite{Chen2023highercup}
\begin{align}
    \sigma(B=d\chi, C=0) = (-1)^{\int\chi\cup d\chi}~.
\end{align}
Comparing with the expression in \eqref{eq:w2}, one can see that $\sigma(B=d\chi, C=0)$ corresponds to the Grassmann integral with $w_2$ fixed to be zero.

\item 
We then define $\sigma(B,C)$ with generic exact $C=d\lambda$. It is given in the form of
\begin{align}
    \sigma(B,C=d\lambda) = \int\prod_f d\vartheta_f d \overline\vartheta_f\prod_{\Delta} u(\Delta)~,
\end{align}
with
\begin{align}
    u(\Delta) = \theta_{\hat 1_-}^{B(\hat 1_-)}\theta_{\hat 2_+}^{B(\hat 2_+)}\theta_{\hat 3_-}^{B(\hat 3_-)} 
    \overline\theta_{\hat 1_+}^{B(\hat 1_+)}\overline\theta_{\hat 2_-}^{B(\hat 2_-)}\overline\theta_{\hat 3_+}^{B(\hat 3_+)} (c_\Delta^\dagger)^{\frac{d\hat C}{2}(\Delta)}~,
\end{align}
which is Grassmann even due to $dB=d\hat C/2$. The integral measure $d\vartheta_f d \overline\vartheta_f$ is given by
\begin{align}
    d\vartheta_f d \overline\vartheta_f = 
    \begin{cases}
        d\theta_f d \overline\theta_f \quad  \text{when $\int \hat{f}\cup \lambda = 0$~, } \\ 
         d \overline\theta_f d\theta_f \quad  \text{when $\int \hat{f}\cup \lambda = 1$~, }
    \end{cases}
\end{align}
where $\hat{f}\in C^2(\hat{\mathcal{T}},\Z_2)$ is a 2-cochain which becomes 1 on a single plaquette $f$, otherwise 0. This modification of the integral measure has an effect of fixing $w_2$ to be $w_2 = C=d\lambda$; for instance, when $d\hat C/2=0$, $\sigma(B=d\chi,C=d\lambda)$ is evaluated as
\begin{align}
    \sigma(B=d\chi, C=d\lambda) = (-1)^{\int\chi\cup d\chi + \chi\cup C}~.
    \label{eq: sigma(B,C) when C=d lambda}
\end{align}
Comparing with the expression in \eqref{eq:w2}, one can see that $\sigma(B=d\chi, C=0)$ corresponds to the Grassmann integral with $w_2=C$.
\end{itemize}

\subsection{Gauge invariance of ground state wave function}

We need to check that the ground state wave function introduced in the main text
\begin{align}
    \ket{\Psi} = \sum_{C\in Z^2(\hat{\mathcal{T}},\Z_2)}Z_{\text{WW}}(B,C)\ket{C}_{\text{edges}} \otimes \sigma(B,C)\ket{0}_{\text{F}}~,
\end{align}
is invariant under gauge transformations of $B\to B + d\chi$, therefore has a well-defined expression. 
The operator $\sigma(B,C)$ described in Sec.~\ref{subsec:sigma(B,C)} shifts under gauge transformations as 
\begin{align}
    \sigma(B+d\chi,C) = \sigma(B,C)(-1)^{\int B\cup_1 d\chi + \chi\cup d\chi+\chi\cup C} = \sigma(B,C)(-1)^{\int \chi\cup B + B\cup \chi + \chi\cup d\chi + dB\cup_1\chi + \chi\cup C }~,
\end{align}
where we used the quadratic property of the Grassmann integral, and that $\sigma(B=d\chi)=(-1)^{\int \chi\cup d\chi +\chi\cup C}$ as described in \eqref{eq: sigma(B,C) when C=d lambda}.

\change{The above phase ambiguity is exactly identical to that of the Walker-Wang anyon diagram $Z_{\text{WW}}(B,C)$ under moving pink $\psi$ lines: the same phase ambiguity appears according to the framing anomaly of the fermionic line~\cite{Kapustin2017bosonization} and mutual braiding of $\psi$ and $v$, which corresponds to the 't Hooft anomaly of the $\Z_2$ 1-form symmetry with the response $\pi (Sq^2B + B\cup C) = \pi (B\cup B + dB\cup_1 B+ B\cup C)$.}

\change{This can be evaluated as follows: when $B$ is closed $dB=0$, the phase ambiguity can be directly computed from evaluating $Z_{\text{WW}}(B,C)$. The wave function involving $B$ is given by $(-1)^{\int \phi\cup B + \phi\cup C}$ using a 1-cochain $\phi$ satisfying $d\phi = B$, where $(-1)^{\int \phi\cup B}$ is the self-linking number encoding fermionic statistics of the $\psi$ line, and $(-1)^{\int \phi\cup C}$ encodes mutual statistics between $v,\psi$.
Then the gauge transformations $B\to B+d\chi, \phi\to \phi+\chi$ shift the wave function by $(-1)^{\int \chi B + B\chi + \chi d\chi + \chi C}$ as desired.}

\change{When $B$ is not closed, the pink $\psi$ line of the 3D Walker-Wang wave function is understood as a boundary condition of the $\{1,\psi\}$ Walker-Wang model on a 4D hypercubic lattice: it consists of a $\psi$ line inserted at the Poincar\'e dual of the $\Z_2$ 3-form $\tilde{B}$ satisfying $dB=\tilde{B}$ at the boundary. The phase ambiguity for the $\psi$ framing anomaly of the bulk 4D wave function is given by the 5D response $Sq^2\tilde{B} = \tilde B\cup_1 \tilde B$~\cite{Chen2023highercup}. Therefore the ambiguity on the 3D boundary is given by the trivialization of the bulk response, given by $Sq^2B = B\cup B + dB\cup_1 B$ so that $d(Sq^2B) = Sq^2\tilde B$. Hence the boundary gauge transformation $B\to B+d\chi, \tilde B\to \tilde B$ transforms the wave function by $(-1)^{\int \chi B + B \chi + \chi d\chi + dB\cup_1\chi}$, according to the inflow of the response. There is also a phase ambiguity from the mutual braiding between $v,\psi$ lines, which gives $(-1)^{\int \chi \cup C}$ by the gauge transformation $B\to B+d\chi$. The whole phase ambiguity is then identical to that of $\sigma(B,C)$ as desired.}

Intuitively, the Grassmann integral accounts for the worldline of the local fermion $f$ so that the $B$ corresponds to the diagonal boson $f\psi$ without framing anomaly, and that the wave function $\ket{\Psi}$ becomes invariant under changing the framing of the worldline for $f\psi$ through the gauge transformation of $B$.

\subsection{Detailed description of Hamiltonian terms}

Now we describe the local Hamiltonians of the Walker-Wang type lattice model introduced in the main text,
\begin{align}
    H_{\nu=2} = -\sum_v \mathcal{P}_v - \sum_p \mathcal{P}_p~.
\end{align}
The definitions of these operators are similar to those for the Levin-Wen string-net models~\cite{Levin2005stringnet}.
First of all, the term $\mathcal{P}_v$ on each (black or pink) fusion vertex is a projector that enforces the admissible fusion diagram on each vertex. On a pink vertex with a complex fermion, $\mathcal{P}_v$ also involves a projector onto the desired fermion occupation at each vertex. Namely, if a pair of short edges ending at the pink vertex has anyons $a,b\in\{1,v,\psi,v\psi\}$, then $a\times b^{-1}$ must fuse into 1 or $\psi$, and if it fuses into $a\times b^{-1} = \psi$, then we put an occupied fermion $c^\dagger \ket{0}$ at the pink vertex. $\mathcal{P}_v$ on a pink vertex enforces both fusion constraints on short edges together with the fermion occupations.

\begin{figure}[htb]
\centering
\IfFileExists{Hamiltonian.pdf}{
\includegraphics[width=0.8\linewidth]{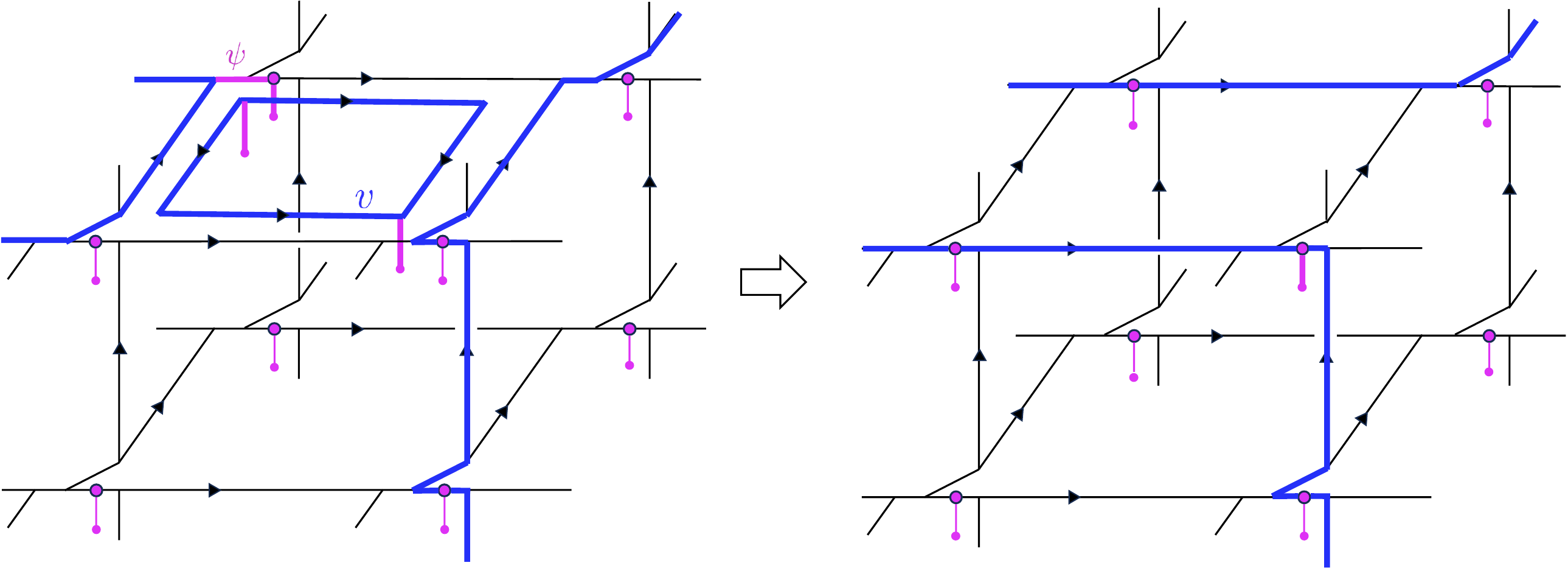}
}{\fbox{\parbox{0.45\linewidth}{\centering Placeholder for \texttt{Hamiltonian.pdf}}}}
\caption{The operator $X_p$ is defined by fusing the small loop formed by $v, v^{-1}$ at the plaquette $p$ into the anyon diagram.
The blue lines denote the $v$ lines. 
A pair of $v$ lines can fuse into a pink $\psi$ line; $\psi$ lines are supported along thickened pink lines. To close up this small loop diagram, $\psi$ lines stemming from vertices of $\partial p$ are connected along the cubic lattice edges according to the $B, B'$ fields. 
The fusion of this diagram is associated with annihilation or creation of fermion operators $c,c^\dagger$, which are accounted for by $F|_{C_p}$. }
\label{fig:Hamiltonian}
\end{figure}

The term $\mathcal{P}_p$ is defined on each plaquette $p$ of the cubic lattice. Depending on the anyon configuration around the plaquette, the term $\mathcal{P}_p$ shifts the anyon labels by acting the small loop that consists of $v, v^{-1}$ along the long edges of $\partial p$. To close up the loop diagram along $\partial p$, a pair of $v$ lines along the loop generally fuses into a pink line that \change{will be connected along cubic lattice edges} (see Fig.~\ref{fig:Hamiltonian}).  Since the complex fermions must be created or annihilated according to the shift of anyon diagrams, it is associated with fermion operators as well.
This has the form of
\begin{align}
    \mathcal{P}_p = \sum_{C|_p}F_{C|_p}\cdot  X_{C|_p}\Pi_{C|_p}~.
\end{align}
Let us explain each operator in the expression. First, $C|_p$ denotes the configuration of $C$ at long edges at a ``neighborhood'' of the plaquette $p$: here the neighborhood is a union of long edges adjacent to four vertices at $\partial p$ (it is a set of 20 edges). The configuration of $C$ on this set of edges is denoted by $C|_p$, and it determines the anyon diagram in the vicinity of the plaquette $p$. Then each operator is described as follows:
\begin{itemize}
    \item $\Pi_{C|_p}$ is a projector onto the states of anyon diagrams in the vicinity of $p$, according to the configuration $C|_p$. 
    \item $X_p$ is an operator that shifts the anyon diagram along the small loop $\partial p$. We want to shift the diagram along the long edges of $\partial p$, by shifting 1 to $v$ and vice versa.
    Depending on the anyon configuration $C|_p$, we prepare a small loop that consists of $v, v^{-1}$ along $\partial p$, such that fusing the loop into the initial diagram results in the desired final diagram. To close up the small loop along $\partial p$, 
    a pair of $v$ lines within the loop generally fuses into a pink $\psi$ line. See Fig.~\ref{fig:Hamiltonian}.
    Fusing this loop to the initial diagram $C|_p$ gives the final diagram $(C+d\hat p)|_p$, where $\hat p$ is a $\Z_2$ 1-cochain of the dual cubic lattice which is 1 at the plaquette $p$, otherwise zero. 
    Fusing the loop to the initial diagram $C|_p$ gives the phase factor $\Theta_{C|_p}$ that consists of a product of $F,R$ symbols. This phase factor $\Theta_{C|_p}$ is a local function of anyon labels on the cubic lattice, hence the operator $X_p$ is local.
    Note that when we define the phase factor $\Theta_{C|_p}$, one needs to fix the initial and final configuration of $B$ fields, to specify the configurations of the $\psi$ lines. Let us write the final configuration of $B$ as $B'$, which will later be used in the expression of $F_{C|_p}$.
    $X_p$ is given in the form of
    \begin{align}
        X_p = \Theta_{C|_p}\prod_e X_e~,
    \end{align}
    where $e$ are short/long edges of the anyon diagram acted by the loop, and $X_e$ is a clock operator of a qudit on the edge that transforms the anyon label accordingly. 
    \item $F_{C|_p}$ is a fermionic operator that consists of a product of complex fermion operators $c^\dagger$, $c$ at vertices of $p$. This operator is given by the comparison of the Grassmann integral operator $\sigma(B,C)$ between the initial and the final configuration of $B,C$. It is defined through the following equation:
    \begin{align}
    F_{C|_p}\cdot \sigma(B,C) = \sigma(B',C+d\hat{p})~.
    \end{align}
    Then $F|_{C|_p}$ is a local operator whose dependence on $C$ becomes local, and involves the creation and annihilation operators of fermions along $\partial p$. 
\end{itemize}

\subsection{Surface topological order}

We obtain the gapped boundary of this model simply by locating the cubic lattice on the Euclidean lattice and truncating the lattice to the cube of finite size, $-L\le x,y,z\le L$. The ground state wave function is given by eliminating Grassmann variables and anyon diagrams outside of the cube region, i.e., setting $B=C=0$ on edges outside of the cube.

In the presence of the boundary, fusing the $v$ lines at the boundary edges gives a line operator $L_v$ for the deconfined $v$ anyon excitation. Its square $L_v\times L_v$ does not act faithfully on the Hilbert space, hence does not braid with other line operators. Therefore the $L_v$ operator indeed generates the $\Z_2$ symmetry. The anyon $v$ has the fusion rule $v\times v=f$ with $f$ a local fermion, since $v^2$ becomes a pink $\psi$ line on the background that terminates with the fermion creation operator $c^\dagger$ at the pink boundary vertex.

\section{The Sixteen-Fold Way}\label{section:MNE}
In this Appendix, we give a short overview of how the $\Z_{16}$ classification for anomalies of 1-form symmetries arises, and discuss its relations with the sixteen-fold way of Kitaev. To see the relation most clearly we will view the sixteen-fold way as the group of minimal nondegenerate extensions of the category $\mathbf{SVect}$. The categories which generate this group are the $Spin(n)_1$ theories \cite{Kitaev_2006, Bruillard2017}, where $n$ is an integer mod 16.

In the theory of braided fusion categories, there is an appropriate notion of a center  of a braided fusion category $\cB$  known as the Müger center, which has objects given by 
\begin{equation}
    Z_2(\cB):=\{x \in \cB |\, \beta_{y,x}\circ \beta_{x,y} = \mathrm{id}_{x\otimes y},\quad  \forall y \in \cB \}
\end{equation}
where $\beta_{x,y}: x\otimes y \to y \otimes x$ is the braiding on $\cB$. A braided category $\cB$ is non-degenerate if $Z_2(\cB) = \mathbf{Vect}$. A nondegenerate extension of $\cB$ is a  choice of fully faithful braided embedding $\cB\hookrightarrow \mathcal{M}$ where $\mathcal{M}$ is a non-degenerate braided fusion category. Upon making the embedding of $\cB$ one can consider the centralizer 
\begin{equation}
    Z_2(\cB\hookrightarrow \mathcal M)\subset \mathcal M\,,
\end{equation}
as the subcategory of all objects in $\cM$ that braid trivially with objects of $\cB$.

\begin{definition}
    A \textit{minimal nondegenerate extension} of a braided category $\cB$ with $Z_2(\cB) = \mathcal{E}$, is a nondegenerate braided fusion category $\cM$ such that $Z_2(\cB \hookrightarrow \cM) = \cE$.
\end{definition}

In \cite{JohnsonFreyd2023MNE} it was shown that a minimal nondegenerate extension of $\cB$ always exists. Essential to the proof was an understanding of (3+1)D TQFTs, and how they can be constructed from the Drinfeld center of a fusion 2-category. The proof goes by establishing an isomorphism $Z(\Mod(\cB))\cong Z(\Mod(\cE))$ of braided fusion 2-categories, from which one can form the non-degenerate braided category $\cM$ with $\cB$ and $\cE$ as centralizing pairs in $\cM$, via tensoring $\cB$ and $\cE$ over the bulk given by $Z(\Mod(\cB))$. Furthermore, the choices of isomorphism for
 $Z(\Mod(\cB))\cong Z(\Mod(\cE))$ correspond to the choices of minimal nondegenerate extension of $\cB$. In the case when $\cE = \mathbf{SVect}$, the relevant (3+1)D TQFT used to study minimal nondegenerate extensions is $\mathcal S= Z(\mathbf{2SVect})$. The choices of minimal nondegenerate extension correspond to the group of automorphisms of this theory. Physically, the automorphisms are implemented by stacking the boundary of $Z(\mathbf{2SVect})$ with some (2+1)D TQFT. It was shown in \cite{GSET} that (3+1)D TQFTs with emergent fermions, which $\mathcal{S}$ falls into, are classified by the degree 5 twisted supercohomology group $SH^{5+\kappa}(B^2\Z_2)$ with $\kappa \in H^2(B^2 \Z_2,\Z_2)$~\footnote{Supercohomology is a generalized  cohomology theory that can be defined using a shifted Pontryagin dual  of the truncation of $ko$ to homotopy groups in degree less than or equal to 2. Namely, we consider $\Sigma^2 I_{\mathbb C^\times}(\tau_{\leq 2} ko)$. 
 In low degrees, twisted supercohomology is equivalent to twisted spin bordism. However the two differ in higher degrees. For the reader that is unfamiliar with supercohomology, it is sufficient to think of it as a way of capturing certain anomalies associated to a classifying space $X$.}. 
 The automorphisms of such theories therefore are classified by  degree four twisted supercohomology $SH^{4+\kappa}(B^2\Z_2) = \Z_{16}$~\cite{johnsonfreyd2020Z2charged, Barkeshli:2023bta, Yang2024gapped}, which is equivalent to the reduced degree four twisted spin bordism group of the space $B^2 \Z_2$. Thus we have related the theories $Spin(n)_1$ with the group of minimal nondegenerate extensions of $\mathbf{SVect}$. This also supports why the $Spin(n)_1$ theories can be used as a starting point to construct fermionic theories that match an anomaly of a given value in $SH^{4+\kappa}(B^2\Z_2)$. 

\section{Characteristic Structures}\label{section:charstructures}
In this appendix we introduce tangential structures for manifolds and how a characteristic structure can give rise to a tangential structure. 

\begin{definition}
    Let $M$ be a manifold. A $\xi$-tangential structure for $M$ is a lift of the map $M \to BO$ of the tangent bundle of $M$ through a prescribed map $\xi: B \to BO$, where $B$ is a general classifying space for some Lie group.
\end{definition}

\begin{example}
 An orientation tangential structure on a manifold $M$ is equivalent to a lift of the classifying map $M\to BO$ through the map $BSO \to BO$. Such a lift exists of the first Stiefel Whitney class of $TM$ vanishes.
\end{example}

Let $M$ be a $n$-dimensional manifold with $\xi$-tangential structure, and consider inserting a topological defect along a submanifold $F\subset M$ with certain properties. When considered as a pair $(M,F)$, one is led to the concept of characteristic pairs:
\begin{definition}
\label{def:charbordism}
Choose an abelian group $A$ and tangential structure $\xi\colon B\to B O$, such that either $A = \Z/2$, or $\xi$ factors through $BSO$. Let $M$ be a manifold with $\xi$-tangential structure, and let $\mathcal P\in H^n(B, A)$. A $(\xi, \mathcal P)$-\textit{characteristic pair} consists of a pair $(M,F)$ where $F$ is a proper submanifold of $M$ Poincaré dual to $\mathcal{P}(M)$ and the boundary of $M$ intersects $F$ precisely and transversely at the boundary of $F$. 
\end{definition}

A manifold $M$ with an embedded submanifold $F$ of the above form is said to have a \textit{characteristic structure} \cite{KT90,Debray:2025iqs}.

The reason why this is interesting is because in many examples the pair $(M,F)$ has a different tangential structure than $M$. The specific characteristic structure we use is discussed in the sequel.

The central tool needed for studying characteristic structures is the Pontryagin-Thom construction, which parametrizes Poincaré dual submanifolds
corresponding to a given cohomology class through the study of maps into a Thom space.

\begin{theorem}[Pontryagin--Thom]\label{construction:PT}
Let $M$ be a closed manifold.
\begin{enumerate}
    \item Let $A$ be a finite abelian group and  $\omega \in H^n(M;A)$. There exists an oriented submanifold $F$ that is Poincaré dual to $\omega$ if and only if the map $\omega\colon M \rightarrow K(A,n)$ lifts across the Thom class map $U\colon M SO_n\to K(A, n)$ to a map $M\rightarrow M SO_n$.
    \item Let $\omega \in H^n(M;\Z/2)$. There exists a (not necessarily oriented) submanifold that is Poincaré dual to $\omega$ if and only if the map $\omega\colon M \rightarrow K(\Z/2,n)$ lifts across the Thom class map $U\colon M O_n\to K(\Z/2, n)$ to a map $M\rightarrow M O_n$.
\end{enumerate}
\end{theorem}

In our construction of the (2+1)D theories at $\nu=1$, it was important that an anyon was inserted along a codimension-2 manifold that is Poincaré dual to $w_2(TM)$. The data of an oriented manifold $M$ along with a submanifold $F$ that is Poincaré dual to $w_2(TM)$ is known as a {Freedman-Kirby characteristic structure}. 

\begin{definition}\label{def:spinc}
    Let $V\to X$ be an oriented vector bundle. A $spin^c$ structure on $V$ is data of a complex line bundle $\mathcal L$ with  $w_2(V) = c_1(\mathcal L) \mod 2 = w_2(\mathcal L)$.
\end{definition}

In light of the Pontryagin-Thom construction, the obstruction to finding a Freedman-Kirby characteristic pair $(M,F)$ is given as an obstruction to the existence of a map $f$ in 

\begin{equation}
    \begin{tikzcd}
        && M SO_2 \arrow[d]\\
        B SO \arrow[rr,"w_2",swap ] \arrow[urr,dotted,"f"] & &K(\Z/2,2)\,.
    \end{tikzcd}
\end{equation}

The pullback square for the space of Freedman--Kirby pairs is therefore given by 
\begin{equation}\label{eq:FKpullback}
    \begin{tikzcd}[column sep=2cm, row sep=2cm]
 B\mathrm{FK} \arrow[r,"L"] \arrow[d,"V"] \arrow[dr, phantom, "\lrcorner", very near start] & M SO_2 \arrow[d, "w_2(L)"] \\
B SO \arrow[r, "w_2(V)"] & K(\Z/2,2)
\end{tikzcd}
\end{equation}
which implies that the data of a FK pair is the data of a rank 2 oriented vector bundle $L$, a map to $B SO$ given by $V$ and an identification $w_2(V) = w_2(L)$. This agrees with Definition \ref{def:spinc}, using the fact that a rank-2 oriented bundle is equivalent to  a complex line bundle.

\begin{proposition}[{\cite[Remark 6.14]{KT90}}]\label{prop:FKpairs}
    The Freedman-Kirby characteristic structure of  is equivalent to a spin$^c$ structure.
\end{proposition}

Given this fact we now patch up a few loose ends in the bulk of the paper around Eq. \eqref{eq:nu1bulk}: 

\begin{enumerate}
    \item The fact that the integral lift of $C$ exists is due to the $spin^c$ nature of the pair $(M,F_{\hat{C}})$.
    \item There is a map $R$ that restricts $(M,F)\rightarrow F$, with the normal bundle $\nu_F\rightarrow F$ remembering how $F$ was embedded in $M$. 
    The tangent bundle of $M$ decomposes  along $F$ as $TF \oplus \nu_F$. The class $w_2(TM)$ restricts to the trivial class on $M\backslash F$, and therefore the restriction of $w_2(TM)$ to $F$ is equal to $w_2(\nu_F)$. This means $w_2(TF)=0$ and $F$ is spin.
    \item The Freedman-Kirby characteristic structures have the property that there is no $\xi$-structure on $M$ that restricts to the $\xi$-structure on $M \backslash F$. If $\xi$ is a spin structure then the surface $F_{\hat{C}}$ is the obstruction for extending a spin structure on $M \backslash F$ to $M$.
\end{enumerate}

The fact that the integral lift of $C$ exists (which is used before Eq. \eqref{eq:nu1bulk}) is a special property of the fact that the 4-manifold $M$ is oriented and we have a  2D submanifold  that is dual with respect to $C=w_2(TM)$, due to the twisted spin structure. In this case, the surface $F_{\hat{C}}$ is naturally a spin manifold, and $F_{\hat{C}}$ obstructs the spin structure on $M\backslash F$ from being equivalent to the one on $M$.


\section{Fermionization and Integral Transforms}\label{section:integraltransforms}
We now reinterpret the map ``Fermionize'' appearing in Equation \eqref{eq:bulkboundary} from a more abstract perspective. In particular, we show that the phase $z_\xi$, familiar from a variety of physical settings \cite{Gaiotto:2015zta, thorngren2019anomaliesbosonization, Kobayashi2019pin,
Tata2022anomalies,Debray:2023iwf}, naturally fits into the general paradigm of integral transforms. This viewpoint connects fermionization to a ubiquitous construction that appears throughout category theory, representation theory, and algebraic geometry.

Consider a span of spaces
\begin{equation}
    \begin{tikzcd}
       & A\times B \arrow[dl,"p_A",swap] \arrow[dr,"p_B"]&\\
        A & &B\,.
    \end{tikzcd}
\end{equation}
The universal property of the product is that giving a map into $A \times B$ is the same as  giving a map $p_A$ into $A$ and a map $p_B$ into $B$. On the category of modules (or sheaves) over the spaces we have the following map from $\Mod_A$ to $\Mod_B$:
\begin{equation}\label{eq:pullpush}
    \begin{tikzcd}
       & \Mod_{A\times B} \arrow[dr,"p_{B*}"]&\\
        \Mod_A  \arrow[ur,"p_A^*",]& & \Mod_B\,.
    \end{tikzcd}
\end{equation}

Consider an object $\cF\in \Mod_A$, and let $K\in \Mod_{A\times B}$ be a kernel object. $K$ defines a functor from $\Mod_A$ to $\Mod_B$ by the formula:
\begin{equation}
    \Phi_K(\cF) := p_{B*}(p^*_A\cF \otimes K)\,,
\end{equation}
i.e. first pulling back using $p^*_A$, tensoring with $K$ and then pushing forward representations using $p_{B*}$\footnote{At this point it is not essential to specify the ambient category that we are taking modules of  these spaces in.}. This concept of a ``pull-push'' map describes many mathematical manipulations and is known in particular cases as an \textit{integral transform}. The name is given by the fact that if one considers working with a function $f(x)$ defined for $x\in A$ and a kernel $K(x,y)$, then an ordinary integral transform with respect to $K(x,y)$ is 
\begin{equation}
    T_K(f)(y) = \int_A K(x,y) f(x) dx\,.
\end{equation}
The ordinary integral transforms admit a natural enhancement from the perspective of categorification, in which functions are replaced by sheaves or modules $\cF$, and kernels define functors between corresponding categories with the pushforward map given by ``integration''. Such constructions play a central role in TQFT, where linear algebra is elevated to higher linear algebra, with state spaces modeled by categories rather than vector spaces. For applications to TQFTs we want a slight variation of the map in Eq. \eqref{eq:pullpush}. We introduce the following notation: let $\mathbf{\Theta}^{\fd}$ be the category of fully dualizable symmetric monoidal $(\infty,n)$-categories. By the Cobordism Hypothesis, $\mathbf{\Theta}^{\fd}$ is the space of $n$-dimensional TQFTs. Furthermore a TQFT with 0-form $G$-symmetry is given by an object $ Z \in \Fun(B G, \mathbf{\Theta}^{\fd})$.

We now apply the integral transform to couple a bosonic theory, with certain symmetries specified, to a fermionic theory with certain symmetries. The example we will eventually apply this to is $Z_{U(1)_4}(B,C)$. Note that we have two $\Z_2$ 1-form symmetries; we will distinguish them as $\Z^B_2$ and $\Z^{C}_2$, with the superscript denoting which background field couples to the symmetry. The classifying space is given by $B^2 (\Z^B_2 \times \Z^C_2)$. We take the target category $\widehat{\bTheta}^{\fd}$ to be the space of (2+1)D spin TQFTs. Such a category has an action by $B \Z^f_2$ where $\Z^f_2$ denotes fermion parity, for which the bosonic theories are the fixed points. In particular the action of $B \Z^f_2$ on $\widehat{\bTheta}^{\fd}$ is given by the following fiber sequence
\begin{equation}
    \begin{tikzcd}
        \widehat{\bTheta}^{\fd}\arrow[r]& \widehat{\bTheta}^{\fd}/\!/ B \Z^f_2 \arrow[d]\\
        & B^2 \Z^f_2\,.
    \end{tikzcd}
\end{equation}
Let $\widehat{X}$ be a \textit{superspace}, i.e. a space $X$ with a map $\kappa: X \to B^2 \Z^f_2$. For our purposes we take such a superspace to be $\widehat{X}=\widehat{B^2 \Z^D_2}$. Let $\Gamma:M\to \widehat{X}$  where $M$ is a spin 3-manifold, such that $\Gamma^*\kappa=D$.
An element of $\Fun(\widehat{B^2 \Z^D_2},\bTheta^{\fd})$ is a (2+1)D TQFT with $\Z_2$ 1-form symmetry whose background field is given by $D$, and the TQFT can couple to twisted spin structure where the twisting is given by $D$.
Given this set up we now consider the following span:
\begin{equation}
    \begin{tikzcd}
       & \Fun(B^2\Z^{B}_2 \times  \widehat{B^2 \Z^C_2}, \widehat{\bTheta}^{\fd} )\arrow[dr,"g_{*}"]&\\
        \Fun(B^2(\Z^{B}_2 \times \Z^{C}_2), \widehat{\bTheta}^{\fd})  \arrow[ur,"f^*",]& & \Fun( \widehat{B^2 \Z^C_2}, \widehat{\bTheta}^{\fd})\,.
    \end{tikzcd}
\end{equation}
Explicitly we can take $Z_{U(1)_4}(B,C)\in \Fun(B^2(\Z^{B}_2 \times \Z^{C}_2), \widehat{\bTheta}^{\fd})$ to be a bosonic theory with some anomaly for $B$ given by $\pi \int B\cup C + Sq^2B$ \footnote{Technically one needs to be careful about the anomaly: it is not quite  given to us by just this setup alone, but extra data that one must provide.}. We pullback along  $f^*$, and tensor by the fermionic theory $z_{\xi}(B)$ also with $\Z_2$ 1-form symmetry with background $B$. We take  $z_{\xi}(B)$ to have the property that it couples to a twisted spin structure $d\xi = w_2 +C$ and has anomaly $\pi\int B\cup C+Sq^2B$. The composite $Z_{U(1)_4}(B,C) \otimes z_\xi(B)$ therefore trivializes the anomaly for $B$ and the pushforward along $g_*$ integrates over $B^2\Z^{B}_2$ i.e. gauges the symmetry. This produces the theory $Z(C,\xi)\in \Fun(\widehat{B^2 \Z^C_2}, \widehat{\bTheta}^{\fd})$ which is a fermionic theory with $\Z_2$ 1-form symmetry and a spin structure twisted by $C$.

\vfill



\end{document}